\documentclass[aps,prl,twocolumn,superscriptaddress,floats]{revtex4-1}
\usepackage{txfonts}
\usepackage{amssymb}
\usepackage{graphicx}
\begin{document} \hbadness=10000
\topmargin -0.8cm\oddsidemargin = -0.7cm\evensidemargin = -0.7cm
\title{Novel bonding patterns and optoelectronic properties of the two-dimensional Si$_x$C$_y$ monolayers}
\author{Dong Fan}
\affiliation{College of Materials Science and Engineering, Zhejiang University of Technology, Hangzhou 310014, China}
\author{Shaohua Lu}
\email{lsh@zjut.edu.cn}
\affiliation{College of Materials Science and Engineering, Zhejiang University of Technology, Hangzhou 310014, China}
\author{Yundong Guo}
\affiliation{School of Engineering and Technology, Neijiang Normal University, Neijiang, 641000, China.}
\author{Xiaojun Hu}
\email{huxj@zjut.edu.cn}
\affiliation{College of Materials Science and Engineering, Zhejiang University of Technology, Hangzhou 310014, China}

\date{\today}
\begin{abstract}
The search of new two-dimensional (2D) materials with novel optical and electronic properties is always desirable for material development. Here, we report a comprehensive theoretical prediction of 2D SiC compounds with different stoichiometries from C-rich to Si-rich. Besides the previously known hexagonal SiC sheet, we identified two types of hitherto-unknown structural motifs with distinctive bonding features. The first type of 2D SiC monolayer, including \emph{t}-SiC and \emph{t}-Si$_2$C sheet, can be described by tetragonal lattice. Among them, \emph{t}-SiC monolayer sheet is featured by each carbon atom binds with four neighboring silicon atoms in almost the same plane, constituting a quasi-planar four-coordinated rectangular moiety. More interestingly, our calculations demonstrate that this structure exhibits a strain-dependent insulator-semimetal transition, suggesting promising applications in strain-dependent optoelectronic sensors. The second type of 2D SiC sheet is featured by silagraphyne with acetylenic linkages(-C$\equiv$C-). Silagraphyne shows both high pore sizes and Poisson's ratio. These properties make them a potentially important material for applications in separation membranes and catalysis. Moreover, one of the proposed structures, $\gamma$-silagraphyne, is a direct-band-gap semiconductor with a bandgap of 0.89 eV, which has a strong absorption peak in the visible-light region, giving a promising application in ultra-thin transistors, optical sensor devices and solar cell devices.
\end{abstract}
\maketitle
Since the demonstration of the first isolated graphene sheet in 2004, 2D atomic crystals have received much attention. For graphene, due to its many extraordinary properties, it has potential applications in a wide range of areas. However, the pristine graphene is a gapless semi-metal, which means that it is difficult to control the number of carriers. This dramatically limits its applications in the field effect transistor, photovoltaic cell, and etc. Thus, the subject of finding new 2D materials beyond graphene is one of the most active fields of current material research. These research include graphyne, single-layer hexagonal boron nitride (\emph{h}-BN),\cite{watanabe2004} transition metal oxides or chalcogenides,\cite{he2012layered},\cite{wang2015physical} phosphorene,\cite{li2014} as well as group IV−VI and III−V layered crystals.\cite{csahin2009} Particularly, besides graphene and graphyne, a strong research topic in group-IV 2D elemental monolayers have sprung up in recent years. However, these group-IV 2D elemental derivatives show the properties of Dirac fermion behavior without spin-orbit coupling, which create a set of challenges for application in conventional electronic devices due to the lack of band gap at the Fermi level.

2D SiC have recently emerged as a promising material with tunable band gaps for potential applications in optoelectronics and electronics.11 Especially, inspired by the successful syntheses of the graphene-like hexagonal SiC sheet in experiment,12 a few carbon-rich SiC monolayers, such as \emph{para}-SiC$_3$,\cite{ding2014} \emph{g}-SiC$_2$,\cite{zhou2013} and \emph{pt}-SiC$_2$,\cite{li2010} were predicted at one particular stoichiometry. Among all of these newly structures, \emph{g}-SiC$_2$ sheet, a direct band gap of 1.09 eV is nearly ideal material for flexible optoelectronic devices. Recently, using first-principle calculations coupled with the cluster expansion method, Shi et al. reported a structural search on SiC sheets with different stoichiometry.\cite{shi2015} However, their structure search is limited to graphene-like hexagonal lattice. Therefore, in order to further explore the configuration space and unique properties of SiC sheets, a global structure search with both variable lattice type and stoichiometry is required. 

In this work, we report a range of low-energy Si$_x$C$_y$ monolayers with distinguished bonding patterns and electronic properties using unbiased particle swarm optimization (PSO) structure search algorithm.\cite{wang2010} The flat hexagonal lattices have been successfully predicted as the low-energy structures for previously reported \emph{g}-SiC$_2$, SiC and many other Si$_x$C$_y$ monolayers in this work. Based on the comprehensive analysis and simulations, we find 2D tetragonal silagraphene named \emph{t}-SiC and \emph{t}-Si$_2$C sheets with both thermodynamic and mechanical stability. The \emph{t}-SiC exists a buckled tetragonal atomic arrangement formed by strong C2p-Si2p bonding with an out-of-plane electron delocalization. More importantly, a phase transition from direct band gap semiconductor to semi-metal is observed when biaxial tensile strain is exerted on \emph{t}-SiC monolayer, suggesting promising applications in strain-dependent optoelectronic sensors. Moreover, six different silagraphyne sheets with higher pore sizes and Poisson's ratio are found, making them potential candidates for separation membranes and catalytic materials. Especially for $\gamma$-silagraphyne, it exhibits direct band gap and obvious optical absorption in visible-light spectrum, indicating a desirable material in microelectronics, solar cell materials and nanoscale optical sensors devices.

\begin{figure}
 \centering
 \includegraphics[width=1\linewidth,clip=] {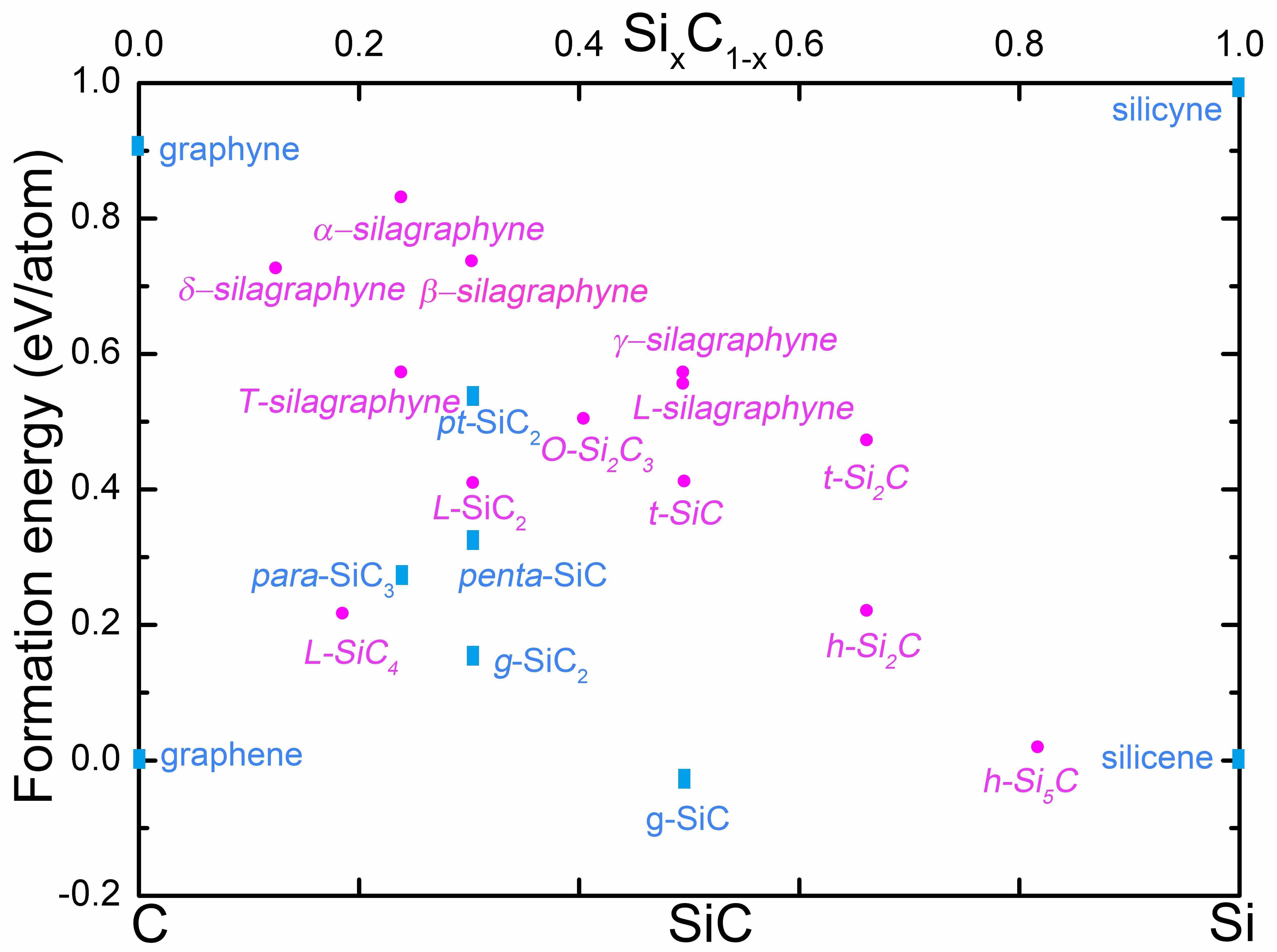}
 \caption{Schematic diagram of the calculated formation energies of Si$_x$C$_y$ sheets under conditions from C-rich to Si-rich. Previously known \emph{para}-SiC$_3$,\cite{ding2014} \emph{g}-SiC$_2$,\cite{zhou2013} \emph{pt}-SiC$_2$\cite{li2010} and \emph{penta}-SiC are also presented for comparison.}\label{fig1}
 \end{figure}

Candidate structures were obtained by global structural optimization method as implemented in CALYPSO code.\cite{wang2010} It has been successfully applied to various crystal surfaces and low dimensional materials.\cite{lu2014,luo2011,yang2015} Structure search was performed with the concentration of carbon ranges from 0.167 to 0.833. The subsequent structural relaxation and total energy calculations were carried out using the density functional theory (DFT) in applying general gradient approximation (GGA) in the Perdew-Burke-Ernzerhof (PBE)\cite{perdew1996} parameterization for exchange correlation potential as implemented in the Vienna \emph{ab initio} simulation package (VASP).\cite{kresse1996} The plane wave cut off energy was set to 650 eV. The convergence criterion of self-consistent calculations for ionic relaxations was 10$^{-5}$ eV between two consecutive steps and the atomic positions and unit cells were optimized until the atomic forces were less than 0.01 eV/\r{A} without any symmetry constraints. A slab model containing a 20 \r{A} vacuum region in the normal direction was selected to simulate isolated 2D materials. Z-axis was fixed during structure relaxation. Heyd–Scuseria–Ernzerhof (HSE06) functional\cite{heyd2003} was used to calculate band structures of selected configurations. In order to study the dynamical stability of the proposed structures, we performed phonon calculations and ab initio molecular dynamics (AIMD) simulations. Phonon calculations were conducted by the density functional perturbation theory (DFPT) method using the PHONOPY code.\cite{togo2015} AIMD simulations were carried out in the NVT ensemble with a time step of 1 fs for a total time of 5 ps.

We collect structures with the lowest energies at each stoichiometric composition and calculate the formation energy (E$_f$) to evaluate the relative stabilities of the predicted 2D SiC compounds, which is defined as follows:
\begin{equation}
 E_{f}=(E_{SixCy}-x\mu_{Si}-y\mu_{C})/(x+y)
\end{equation}

where E$_f$ denotes the formation energy of the corresponding 2D SiC compounds, E$_S$$_i$, E$_C$ and E$_S$$_i$$_x$C$_y$ are the total energy of a single Si atom, a single C atom, and the 2D Si$_x$C$_y$ compound.

Fig 1 shows the schematic diagram of the calculated formation energies of Si$_x$C$_y$ sheets under conditions from C-rich to Si-rich. Previously known sheets, such as hexagonal SiC, \emph{para}-SiC$_3$, \emph{g}-SiC$_2$, and \emph{pt}-SiC$_2$ have been successfully reproduced in this work, and our calculated formation energies are consistent with previous results. Besides, two types of hitherto unknown structure motifs are revealed in the global structure search. The predicted low-energy structures of several allotropes with unique bonding patterns and the corresponding lattice parameters are listed in Table S1. All 2D compounds with novel bonding geometries are predicted and plotted in Fig 2, S1 and S2. Moreover, kinetic stability of these 2D sheets is verified by both phonon spectrum calculations and AIMD simulations. Next, we perform a systematic analysis of the chemical bonding, mechanical and electronic properties on the newly found structures.

Fig 2a displays the geometric configuration of \emph{t}-SiC monolayer. Its rectangular lattice constants are a = 3.67 \r{A} and b = 3.21 \r{A}, with a thickness c=1.01 \r{A}. One unit cell of \emph{t}-SiC consists of 2 C atoms and 2 Si atoms, respectively. Especially, all the atoms in the supercell bond to four neighboring atoms. Usually, just like graphene, the majority of 2D sheets have three-fold coordinated bonding moieties. All C and Si atoms in the \emph{t}-SiC sheet are quasi-planar tetracoordinated atoms. To the best of our knowledge, this is the first time to report pure tetracoordinated carbon atoms in 2D covalent materials. The stabilizing mechanisms and bonding characteristics are disclosed by analyzing the deformation electron density data, which is defined as the total electronic density of 2D SiC system excluding the electronic densities of isolated Si and C atoms. Deformation charge density of \emph{t}-SiC monolayer is shown in Fig S3, revealing that some electrons are extracted from Si atoms and delocalized around the four Si-C bonds. This is crucial to stabilizing the quasi-planar tetracoordinated moieties. The buckling of t-SiC monolayer binds Si-C bonds in tetragonal lattice and weakens interaction between two neighboring atoms. In order to gain more insights into chemical bonding in \emph{t}-SiC sheet, we plot the electron localization function (ELF) to determine the extent of covalent character. An isosurface of ELF for \emph{t}-SiC is illustrated in Fig S4, with an isovalue of 0.80 au, indicating the formation of strong Si-C $\sigma$ bonds.

\begin{figure}[htbp]
 \centering
 \includegraphics[width=0.9\linewidth,clip=] {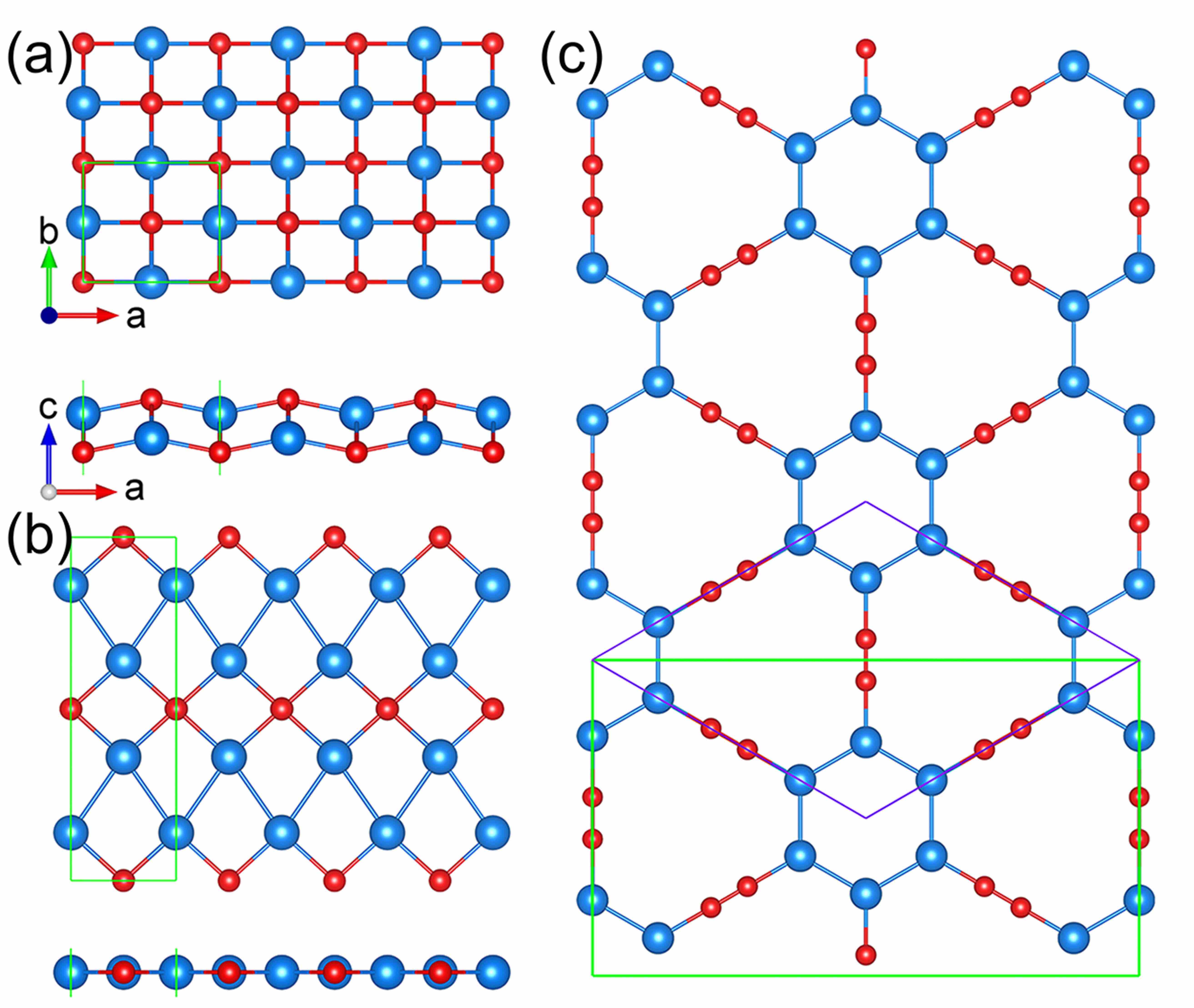}
 \caption{Top (upper) and side (lower) views of the structure of (a) \emph{t}-SiC, (b) \emph{t}-Si$_2$C and (c) $\gamma$-silagraphyne monolayer. The blue and red balls represent Si and C atoms, respectively. The rhombohedra (violet) and rectangular (green) unit cells are presented.}\label{fig2}
 \end{figure}

Another structure (named as \emph{t}-Si$_2$C) with tetragonal lattice is also revealed. As shown in Fig 2b, \emph{t}-Si$_2$C monolayer forms the exactly planar structure and the optimized lattice constants for \emph{t}-Si$_2$C monolayer are a = 2.79 \r{A} and b = 9.09 \r{A}. Fig S3(b) shows the electron transfer from Si atoms to C atoms in \emph{t}-Si$_2$C monolayer. Moreover, carbon 2p electrons are delocalized around the four Si-C bonds in the same plane, suggesting the covalent character of Si-C bonds. For \emph{t}-Si$_2$C, its ELF isosurfaces exhibit two domains: one is the distribution around the centered four Si-C bonds symmetrically, and the other is the Si-Si dimer rows, whereas the bonding between Si atoms is weak compared with Si-C $\sigma$ bonds.

 Besides tetragonal silagraphene, we also demonstrate a new 2D SiC monolayer with acetylenic linkages (-C$\equiv$C-), silagraphyne, namely. Fig 2 and S1 display three 2D SiC compounds, silagraphyne, containing acetylenic linkages that are more stable than graphyne by comparing their formation energies. It is worth noting that the shortest distances between sp-carbon atoms are 1.24 and 1.23 \r{A} of $\alpha$($\beta$)-silagraphyne and $\gamma$-silagraphyne, respectively. This indicates that the localization of electron density is in the binding regions. As demonstrated in Fig S1 and S2, all the structures of silagraphyne, except for L- and $\gamma$-silagraphyne with a buckling of 0.65 and 0.48 \r{A}, are purely planar. It is clear to see that the acetylenic linkages generate pores, which are much larger than those in graphene and graphyne.\cite{puigdollers2016} According to our calculations, $\alpha$-, $\beta$- and $\gamma$-silagraphyne have the inverse of the specific surface area of 0.362, 0.432 and 0.537 mg/m$^2$, respectively. This size of the pore is very important in the desalination of sea water,\cite{xue2013} photocatalyst, lithium-ion battery and energy storage,\cite{li2014} because they have high porosity and high specific surface area compared to graphyne (i.e., 0.379 mg/m$^2$ for $\alpha$-graphyne, 0.461 mg/m$^2$ for $\beta$-graphyne and 0.582 mg/m$^2$ for $\gamma$-graphyne).\cite{puigdollers2016} The calculative ELF maps of silagraphyne monolayers are demonstrated in Fig S5. It's clear to see that all of the ELFs of C-C and Si-C bonds are localized at the bond center, similar to the ELF maps in \emph{t}-SiC monolayer.

The stability of the 2D Si$_x$C$_y$ sheet can first be understood by calculating its formation and cohesive energies, as shown in Fig 1 and Table S1. The results show that newly predicted \emph{t}-SiC is higher in energy than graphene by 0.39 eV/atom, but lower than that of previously reported \emph{pt}-SiC$_2$ by 0.19 eV/atom. Simultaneously, the cohesive energy of \emph{t}-SiC and \emph{t}-Si$_2$C is 5.55 and 4.85 eV/atom, respectively, which are larger than that of silicene (3.93 eV/atom) and lower than that of graphene (7.94 eV/atom). Thus, it is proved that 2D tetragonal SiC monolayer has higher structural stability and stronger binding energy comparative with silicene. Obviously, the silagraphyne monolayers are metastable while they all have lower formation energies with respect to graphene and silicene. In particular, we also note that silagraphyne monolayers are energetically preferable over experimentally confirmed graphdiyne films.\cite{li2010} This implies that the newly found silagraphyne sheets might be synthesized in the experiment. Moreover, we calculated the cohesive energy to evaluate the stability of proposed silagraphyne structures. As shown in Table S1, the cohesive energies of silagraphyne monolayers are ranging from 4.50 to 6.55 eV/atom, which are significantly lower than that of graphene, but still higher than that of silicene, attesting that the silagraphyne sheets are energetically stable.

\begin{figure}[tbhp]
 \centering
 \includegraphics[width=1.0\linewidth,clip=] {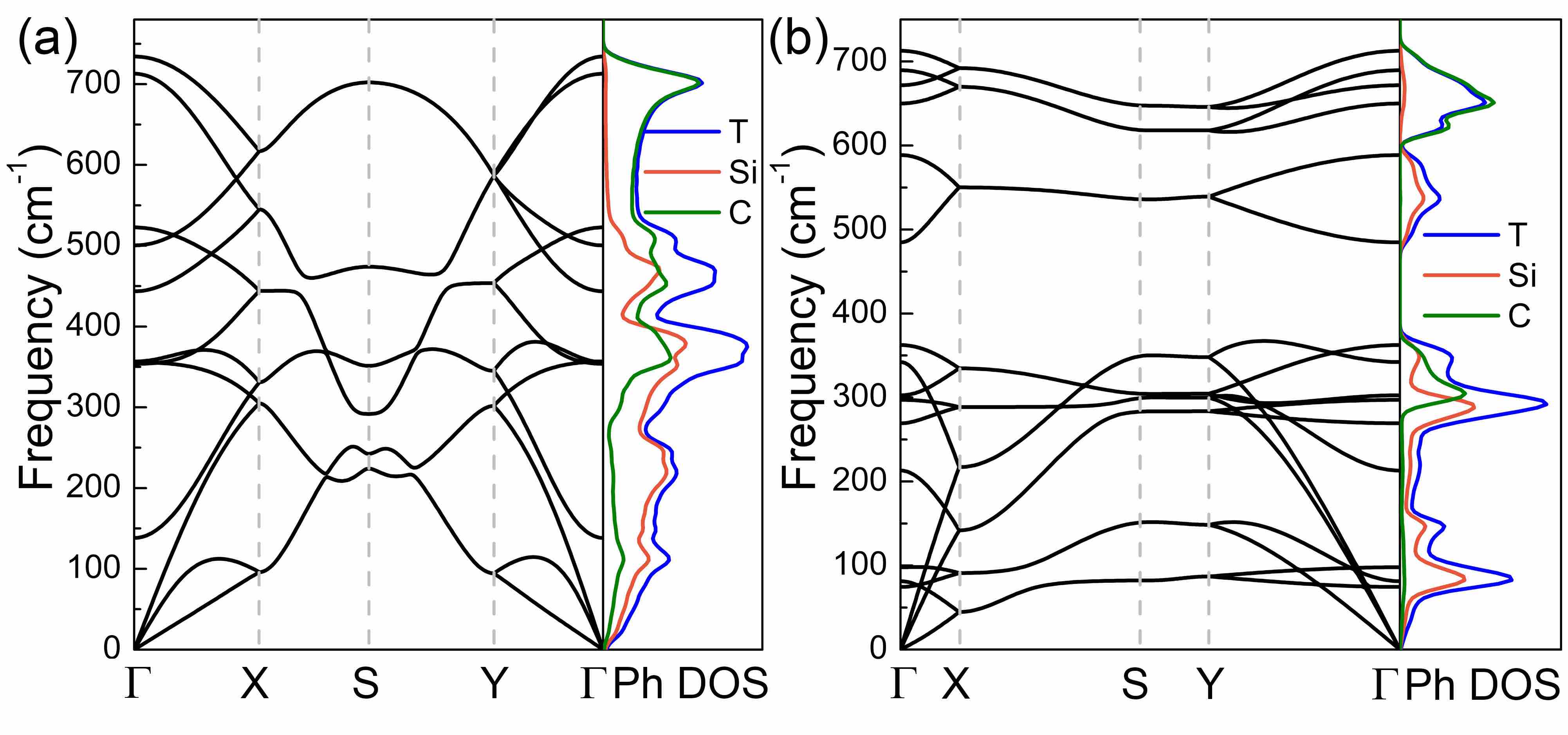}
 \caption{Phonon band dispersions (left panel) and partial PhDOS (right panel) of (a) \emph{t}-SiC and (b) \emph{t}-Si C monolayer are calculated by linear response theory. $\Gamma$(0, 0, 0), X(1/2, 0, 0), S(1/2, 1/2, 0) and Y(0, 1/2, 0) refer to special points in the first Brillouin zone of reciprocal space. There is no imaginary frequency in the whole path.}\label{fig3}
 \end{figure}
To examine the dynamic stability of proposed structures, the phonon dispersion spectra and phonon density of states (Ph DOS) were calculated by using DFPT method along the high-symmetry lines, as shown in Figure 3 and Figure S6. It is observed that there is no imaginary phonon frequency in the entire Brillouin zone, confirming that tetragonal silagraphene sheets are dynamically stable. For \emph{t}-SiC and \emph{t}-Si$_2$C monolayer, we note that the highest frequency reaches up to 735 and 713 cm$^{-1}$, respectively, which is slightly higher than that of silicene (580 cm$^{-1}$)\cite{cahangirov2009} and MoS$_2$ monolayer (473 cm$^{-1}$),\cite{molina2011} indicating robust Si-C bonds in these 2D systems. Analysis of partial PhDOS reveals that the highest frequency of \emph{t}-SiC is mainly originated from C atoms. The phonon dispersion curves of \emph{t}-Si$_2$C indicate the presence of phonon gap and have the separation of acoustic and optical branches with the maximum vibrational frequency at $\sim$ 368 cm$^{-1}$. Detailed analysis of the PhDOS reveals that the lower acoustic modes are associated with the constituent of Si atoms, while the high frequency vibrational modes attribute to the Si-C stretching modes. To further understand the stability of these unique structures at ambient conditions, we have performed AIMD simulations within the canonical NVT ensemble at 1000 K for 5 ps with a time step of 1 fs. Total energy development of simulation time and snapshots taken at the end of each time is shown in Figure S7-S9. During the final NVT simulation, \emph{t}-SiC and \emph{t}-Si$_2$C sheets are stable up to a temperature of 800 K, while other silagraphene and silagraphyne can maintain the structural integrity at 1000 K (Figure S8, S9). The above results reveal that these 2D SiC monolayers exhibit a high thermal stability.

For a mechanically stable 2D material, the elastic constants just have to fulfill the following requirements:\cite{andrew2012} C$_{11}$C$_{22}$- C$_{12}$C$_{21}$ $>$ 0 and C$_{66}$ $>$ 0. For silagraphene, the calculated elastic constants satisfy this formula, and the calculated C$_{66}$ is positive (Table S2), indicating that all the proposed silagraphene sheets are mechanically stable. The in-plane stiffness modulus, which can be derived from the elastic constants by: E = (C$_{11}$C$_{22}$ - C$_{12}$ C$_{21}$)/C$_{11}$ . Tetragonal silagraphene sheets present elastic in-plane stiffness with 119 and 169 GPa$\cdot$nm along the \textbf{b} direction for \emph{t}-SiC and \emph{t}-Si$_2$C, respectively, which are more than double of that of silicene. The calculated elastic constants of six silagraphyne monolayers also satisfy the mechanical stability criteria, indicating that the 2D silagraphyne monolayers are mechanically stable (Table S3). Interestingly, the calculated in-plane stiffness values for three kinds of silagraphyne going from $\alpha$-silagraphyne, to $\beta$- and $\gamma$-silagraphyne are 17.00, 26.33 and 47.55 GPa$\cdot$nm, respectively. The in-plane stiffness increases with decreasing specific surface area and is significantly lower than that of graphyne. Thus, these 2D sheets inserting acetylenic linkages are much softer than graphyne and graphene, due to the lower planar packing densities of these structures with respect to the graphene. Similar to graphyne, $\alpha$-, $\beta$- and $\gamma$-silagraphyne have Poisson’s ratios between 0.5 and 1, suggesting a higher structural deformation along the perpendicular direction to the plane.\cite{puigdollers2016}

Figure 4 shows the band structures (calculated via HSE functional) of \emph{t}-SiC and \emph{t}-SiC structures. As presented in Figure 4a, \emph{t}-SiC is a direct band gap semiconductor with a band gap of 1.84 eV (0.85 eV at the PBE level) at the $\Gamma$ point. Its valence band maximum (VBM) is contributed by the Si-3s ($\sigma$) orbitals, while the conduction band minimum (CBM) is contributed by Si-3p ($\pi$) and C-2p ($\pi$) orbitals (Figure S10). However, different from \emph{t}-SiC, our band structure calculations show that \emph{t}-Si$_2$C is metallic with a finite density of states at the Fermi energy. Figures 4c(e) and 4d(f) indicate the highest occupied orbital (HOMO) and the lowest unoccupied orbital (LUMO) at the $\Gamma$ point, respectively. It is interesting to note that, for \emph{t}-Si$_2$C, the HOMO clearly shows the strong bonding between two Si atoms, while LUMO is mainly distributed around the Si atoms.

Supplementary Figure S11 shows the phonon dispersion for \emph{t}-SiC sheet when biaxial strain is applied, for strain values of 1\%, 9\% and -9\%. There is not any imaginary phonon mode in phonon dispersion, so its strained monolayer is also stable. Subsequently, for \emph{t}-SiC sheet, we calculate the HSE band structure at different biaxial strain (Figure 4g). Notably, we find that the band structure can be transformed from direct band gap to semimetal at a relatively suitable biaxial strain (10\%). The direct band gap persists under a tensile strain of 1 to 4\% and the band gap decreases from 1.72 to 1.31 eV. According to Shockley-Queisser limit,\cite{shockley1961} for single-gap photovoltaic devices with the band gap of 1.34 eV, a maximum photoelectric conversion efficiency of 33.7\% can be achieved. Thus, the direct band structure of small strained monolayer supplies a favorable advantage for the applications in photoelectric devices.
\begin{figure}
 \centering
 \includegraphics[width=1.0\linewidth,clip=] {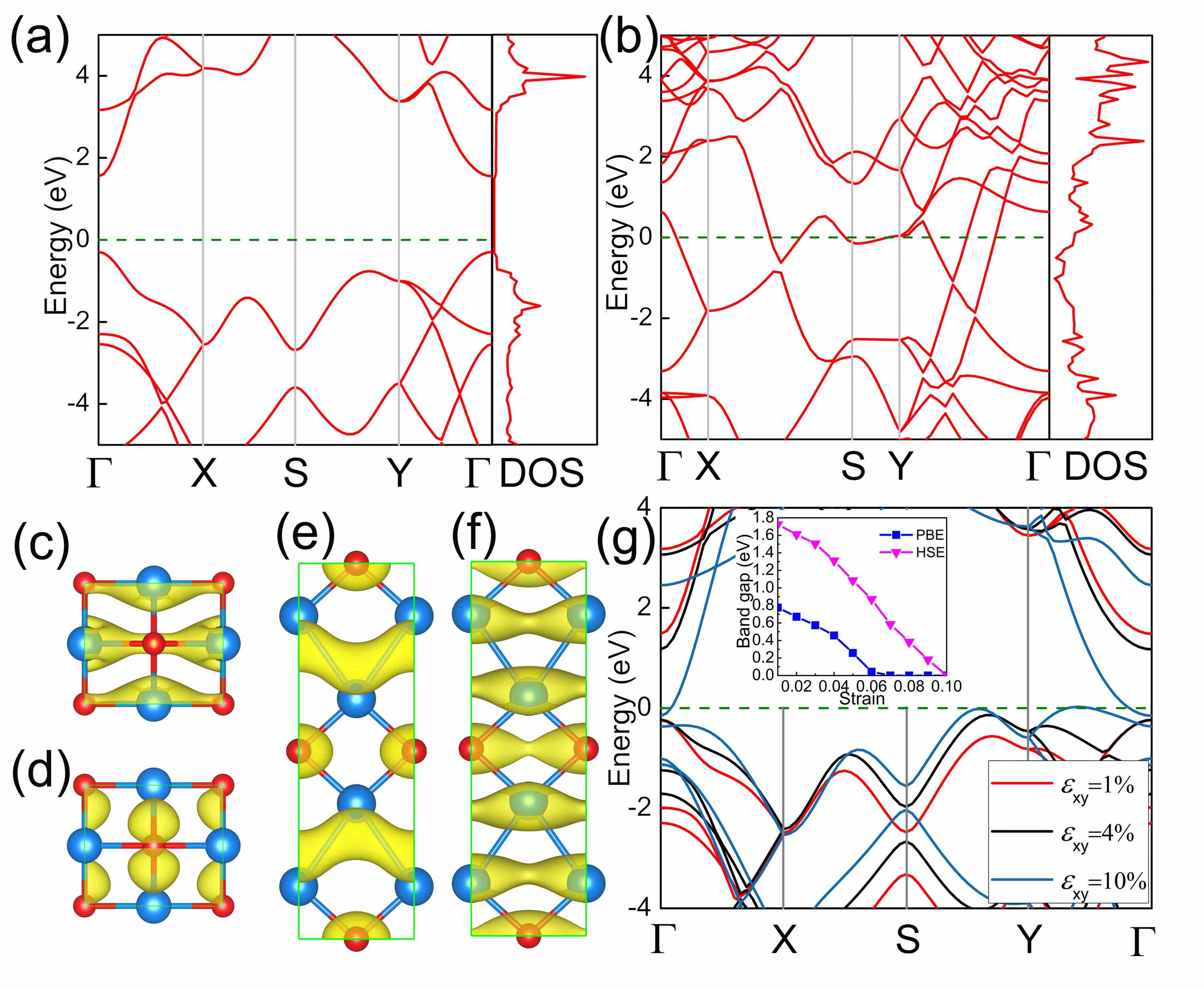}
 \caption{Electronic band structure and DOS of (a) \emph{t}-SiC and (b) \emph{t}-Si$_2$C structures are calculated by using HSE06 functional. Band-decomposed charge density distributions are illustrated in c to f. (c) The highest occupied electronic state and (d) the lowest unoccupied electronic state of \emph{t}-SiC at the $\Gamma$ point. (e) The highest occupied electronic state and (f) the lowest unoccupied electronic state of \emph{t}-Si C at the $\Gamma$ point. The Fermi level is set to zero and marked by the dashed lines. (g) Band structure of \emph{t}-SiC with different biaxial strains. Inset shows the band gap of monolayer as a function of biaxial strains (calculated by PBE and HSE06 functionals).}\label{fig4}
 \end{figure}
Figure 5a and Figure S12 demonstrate the electronic band structures of a few selected silagraphyne monolayer calculated with PBE functional. It is observed that $\alpha$-silagraphyne and $\beta$-silagraphyne show the presence of Dirac points at the Fermi level, but they show different number and location of Dirac points. $\alpha$-silagraphyne has Dirac points located at K high symmetry points, while $\beta$-silagraphyne exists Dirac points in the $\Gamma$-M direction. These interesting results are similar to the electronic structure of graphyne. However, $\gamma$-silagraphyne and T-silagraphyne show direct band gaps, where the CBM and VBM are located at the M and $\Gamma$ points in the Brillouin zone, respectively. Using the HSE functional, we note that $\gamma$-silagraphyne and T-silagraphyne have the band gap of 0.89 and 2.02 eV, respectively, while $\delta$-silagraphyne and L-silagraphyne show semimetal property.

Band gap plays a fundamental role in semiconductor materials, due to the strong dependence of band gap and solar cells energy conversion efficiency. The characteristics of moderate direct band gap for 2D SiC monolayers make it promising solar-cell absorption materials. Thus, optical absorption spectra of predicted low energy structures were calculated at the HSE06 level. For comparison, calculations were also carried out for silicon. Our results for the silicon curves are in accord with previous calculations.\cite{guo2015} Figure 5b shows the optical absorption spectra. According to the spectral range, we separated the absorption spectrum into three parts, namely the infrared, visible, and ultraviolet regions, respectively. Analogous to silicon, due to the large direct band gap, \emph{t}-SiC monolayer can absorb photons in the ultraviolet range mostly. Surprisingly, $\gamma$-silagraphyne can absorb the sunlight at lower energies than that of \emph{t}-SiC sheet, implying that this structure can capture more sunlight and make the efficient use of solar energy compared to silicon.

The aforementioned results suggest that tetragonal silagraphene and silagraphyne are very promising candidates for application in future optoelectronic nanodevices due to their unique structural and electronic properties. Thus, it is desirable to synthesize these new monolayers experimentally. Compared to the honeycomb structure, tetragonal lattice seems impossible according to the valence electron pair repulsion rule.\cite{jolly1984} However, in fact, planar tetracoordinated carbon has been first proposed by Hoffman \emph{et al}. (1970)\cite{hoffmann1970} and later synthesized in metal compound molecules (1977).\cite{cotton1977} In organic chemistry, the compound having the shape of a square pyramid is called [3.3.3.3] fenestrane (or pyramidane), consisting of a central carbon atom which serves as a common vertex for four fused carbocycles. Experimentally, although this chemical compound has not been synthesized yet, the synthesis of related Ge[C$_4$(SiMe$_3$)$_4$] and Sn[C$_4$(SiMe$_3$)$_4$] have been reported.\cite{lee2013} Recent theoretical results also elaborate the existence of planar 4-coordinated C atoms in 2D materials, such as 2D BC compounds,\cite{luo2011} \emph{t}-TiC\cite{zhang2012} and \emph{penta}-graphene\cite{zhang2015} \emph{et al}. Thus, in consideration of the chemical similarity between carbon and silicon, it provides a possibility to produce a newly tetragonal silagraphene experimentally.
\begin{figure}
 \centering
 \includegraphics[width=0.9\linewidth,clip=] {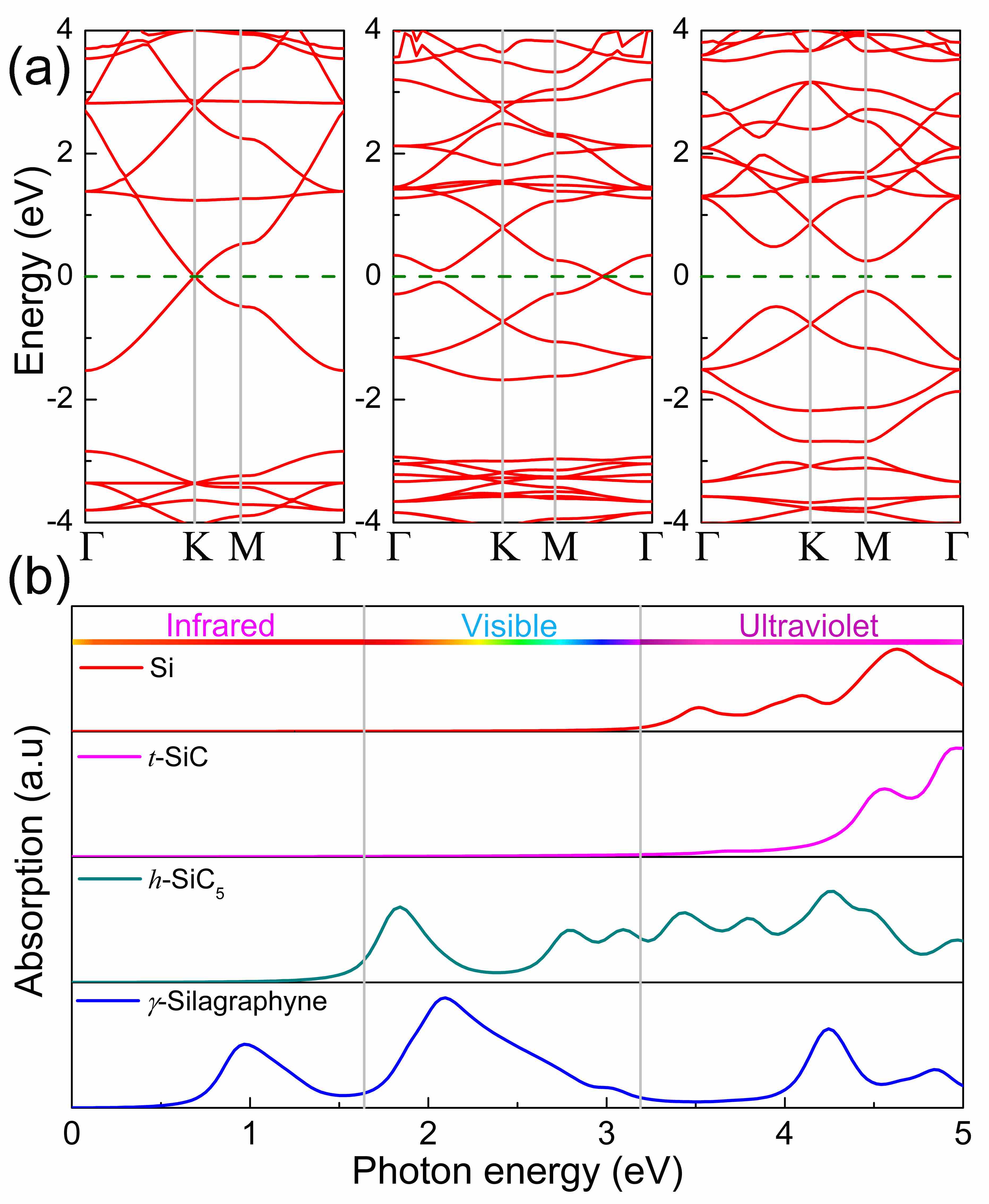}
 \caption{(a) Calculated band structure of $\alpha$-silagraphyne (left), $\beta$-silagraphyne (middle), and $\gamma$-silagraphyne (right) monolayers. The Fermi level is set to zero and marked by dashed lines. (b) Optical absorption coefficient of the proposed structures, silicon data is presented for comparison. (calculated by HSE06 functional).}\label{fig5}
 \end{figure}

Here, for \emph{t}-SiC monolayer, a possible strategy to obtain this sheet by hydrogen intercalation to break the subsurface C-Si covalent bonds in 3C-SiC is proposed. We first isolate \emph{t}-SiC sheet from 3C-SiC substrate (Optimized structural model is shown in Figure 6a) and check its structural stability by phonon calculation (Figure 6b). Next, the effect of hydrogen intercalation in the subsurface of 3C-SiC for stripping a \emph{t}-SiC monolayer is investigated by performing the AIMD simulation shown in Figure 6c. After a period of time, the system has reached thermodynamic equilibrium, a \emph{t}-SiC monolayer with buckled structure is observed and striped from the 3C-SiC film surface. Although this process may have high energy barriers, there is an experimental evidence that hydrogen/deuterium can selectively interact with 3C-SiC substrate at subsurface by using high-resolution electron energy loss spectroscopy and synchrotron radiation based photo-emission spectroscopy techniques.\cite{soukiassian2013}

\begin{figure}
 \centering
 \includegraphics[width=1.0\linewidth,clip=] {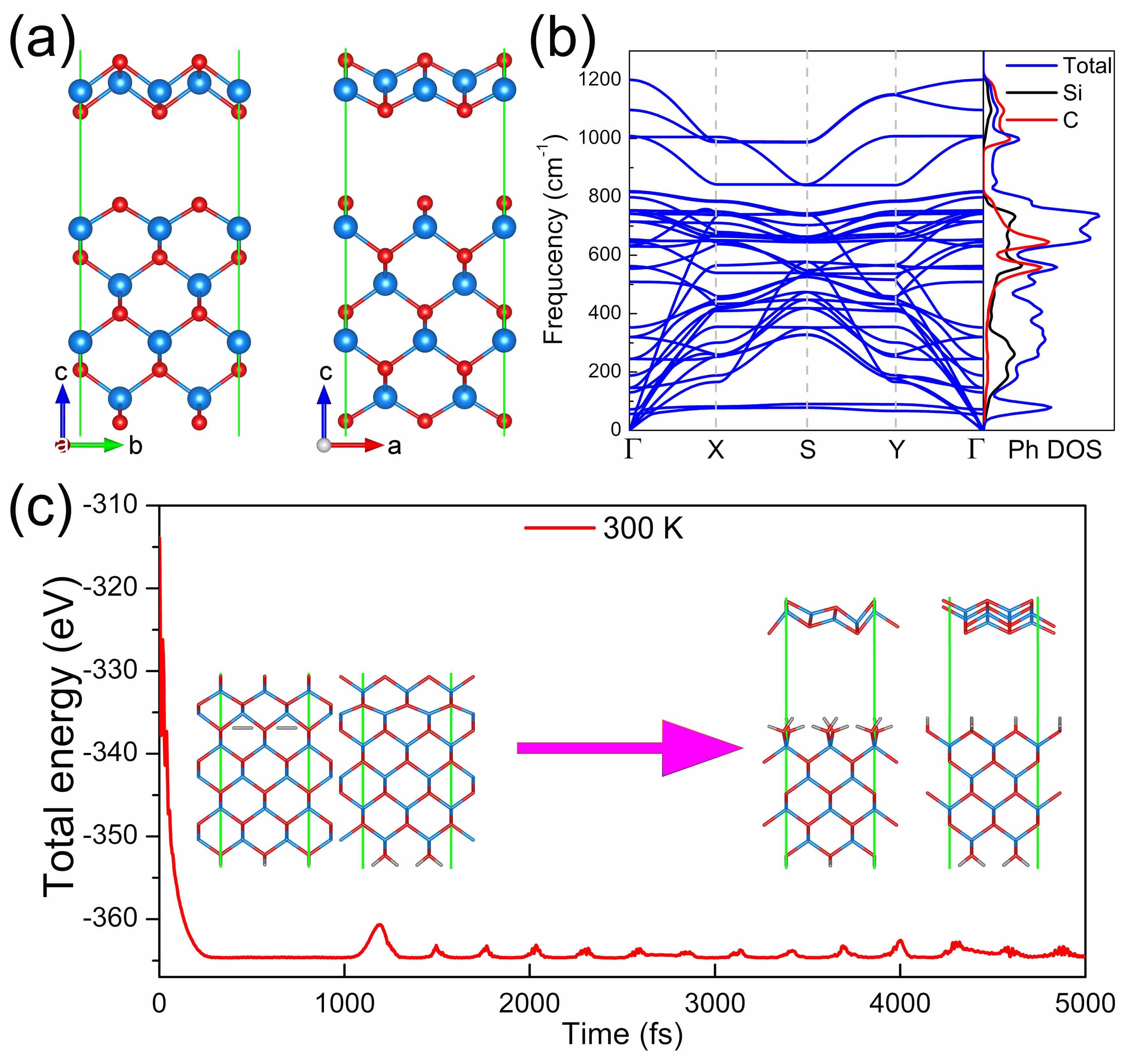}
 \caption{(a) Structure of 3C-SiC (0001) surface containing t-SiC monolayer and corresponding Phonon spectrum (b). (c) The calculated energy evolution of hydrogen in intercalated 3C-SiC surface during AIMD simulation at 300K. The insets show the initial and final structures viewed along a and b direction, respectively.}\label{fig6}
 \end{figure}

In conclusion, we find two types of 2D SiC monolayers with unique bonding structures: tetragonal silagraphene and silagraphyne. The proposed tetragonal silagraphene sheets (\emph{t}-SiC and \emph{t}-SiC) exhibit a novel planar (quasi) tetracoordinated Si-C pattern, while the silagraphyne shows an acetylenic linkages (\emph{sp} carbon atoms) in the 2D crystal lattice. Then, kinetics and dynamics stability of the predicted monolayers were investigated by performing AIMD simulations and the calculations of phonon dispersion curves, respectively. Especially, for \emph{t}-SiC monolayer, each Si atom is coordinated to four C atoms to form a tetragonal motif, which closely resembles pyramidane in structures with an approximate D 2h symmetry. To our knowledge, such monolayer purely consists of quasi-planar covalent tetracoordinated carbon and silicon atoms have never been reported in the previous literature. Owing to its unique atomic patterns and strain-induced semiconductor to metal transition properties, \emph{t}-SiC shows promise for applications in optoelectronic sensors when the suitable biaxial tensile strain is applied. Silagraphyne sheets have higher pore sizes, Poisson's ratio, and tunable band gap energies varying from 0.00 to 2.02 eV. These structural and electronic properties make silagraphyne as alternative materials for special applications that need softer materials and the next-generation photovoltaic devices. Finally, as a theoretical exploration, we propose a fabrication process by using hydrogen intercalation to break the subsurface C-Si covalent bonds in 3C-SiC. In view of distinct bonding structure and dynamics stability to endure the biaxial strain, these novel structures might be synthesized experimentally and used to fabricate photoelectric conversion device with a high efficiency.

This work was supported by the National Natural Science Foundation of China (Grant Nos.50972129, 50602039, and 11504325), and Natural Science Foundation of Zhejiang Province (LQ15A040004). This work was also supported by the international science technology cooperation program of China (2014DFR51160).

\bibliographystyle{unsrt}
\bibliography{ref}

%
%

%
%

\end{document}